\documentclass[a4paper,12pt]{article}
\newcommand{\bea}{\begin{eqnarray}}
\newcommand{\eea}{\end{eqnarray}}
\newcommand{\ba}{\begin{array}}
\newcommand{\ea}{\end{array}}
\newcommand{\half}{\frac{1}{2}}

\newcommand{\be}{\begin{equation}}

\newcommand{\ee}{\end{equation}}
\newcommand{\bt}{\begin{teo}}
\newcommand{\et}{\end{teo}}
\newcommand{\ep}{\epsilon}
\newcommand{\s}{\sigma}
\newcommand{\la}{\lambda}

\newcommand{\als}{\alpha^2}
\newcommand{\bets}{\beta^2}

\begin{document}

\begin{center}

\Large

{\bf Some generic properties of level spacing distributions of 2D real random matrices}\\
\vspace{0.5in}
\large
{\bf Siegfried Grossmann$^1$} and {\bf Marko Robnik$^{2}$}\\
\end{center}

\vspace{0.2in}

\noindent
$^1$Fachbereich Physik der Philipps-Universit\"at, Renthof 
6, D-35032 Marburg, Germany 

\noindent
$^2$CAMTP - Center for Applied Mathematics and Theoretical Physics,
University of Maribor, Krekova 2, SI-2000 Maribor, Slovenia
\\\\
{\bf e-mail:} Grossmann@physik.uni-marburg.de, Robnik@uni-mb.si
\\\\
\noindent
{\bf Abstract:}
We study the level spacing distribution $P(S)$ of 2D real random matrices 
both symmetric as well as general, non-symmetric. In the general case we restrict ourselves 
to Gaussian distributed matrix elements, but different widths of the various matrix elements 
are admitted.  The following results are obtained: An explicit exact formula for $P(S)$ 
is derived and its behaviour close to $S=0$ is studied analytically, showing that there is 
linear level repulsion, unless there are additional constraints for the probability distribution 
of the matrix elements. The constraint of having only positive or only negative but otherwise 
arbitrary non-diagonal elements leads to quadratic level repulsion with logarithmic corrections.
These findings detail and extend our previous results already published in a preceding paper.
For the {\em symmetric} real 2D matrices also other, non-Gaussian statistical distributions are 
considered. In this case  we show for arbitrary statistical distribution of the diagonal 
and non-diagonal elements that the level repulsion exponent $\rho$ is always $\rho = 1$, provided
the distribution function of the matrix elements is regular
at zero value. If the distribution function of the matrix elements 
is a singular (but still integrable) power law near zero value of $S$, the level spacing distribution 
$P(S)$ is a fractional exponent pawer law at small $S$. The tail of $P(S)$ depends on further 
details of the matrix element statistics. We explicitly work out four cases:
the constant (box) distribution, the Cauchy-Lorentz distribution, the exponential distribution
and, as an example for a singular distribution, the power law distribution for $P(S)$ near zero 
value times an exponential tail.

\noindent
Submitted to {\em Zeitschrift f\"ur Naturforschung A} 

\section{Introduction}
\label{introduction}

Random matrix theory \cite{Haake}-\cite{Rob} has important applications 
quite generally in the statistical description of complex systems as 
e.g. for complex nuclei, for which it has been originally developed, or for
chaotic systems with just a few degrees of freedom as treated in
quantum chaos. Usually one restricts oneself to Gaussian ensembles
of random matrices, meaning that the matrix elements have a Gaussian
distribution where the diagonal matrix elements have all the same
dispersion, the off-diagonal elements also have the same dispersion,
but the former is by a factor 2 larger than the latter one.
See the remarks in section 2. This ensemble is
the only one which is invariant under symmetry transformations
of the underlying matrix ${\bf A}$. If ${\bf A}$ is real and symmetric, the group
of relevant transformations consists of the orthogonal transformations 
and we speak of the Gaussian Orthogonal Ensemble (GOE), while
for Hermitian ${\bf A}$ the relevant group of transformations is that of 
the unitary transformations and we speak of the Gaussian Unitary Ensemble
(GUE) of random matrices. The property of the matrix element distribution to be Gaussian is a
direct consequence of just two assumptions, namely statistical
independence of the distributions of the matrix elements and invariance
against the group of appropriate transformations. We usually have in mind infinite 
dimensional matrices, although for many purposes finite dimensionality is useful and sufficient.

Several generalizations are possible. One is the generalization 
towards general non-normal matrices, either fully complex 
\cite{Gin} (see also \cite{Haake}) but still Gaussian (invariant), 
or real positive but no longer invariant against the above group of transformations 
while still Gaussian (with different variances for different matrix elements) \cite{GroRob}.
In the latter paper \cite{GroRob} we have also treated 2D real symmetric matrices whose 
matrix elements are Gaussian distributed but with different variances of the diagonal and the 
non-diagonal matrix elements. Thus they no longer 
enjoy GOE invariance, but still have Gaussian distributed matrix elements.

In the present paper we study the level spacing distribution $P(S)$ of
2D real random matrices by considering general distributions of the
matrix elements, going beyond Gaussian. Therefore these ensembles of random 
matrices are no longer invariant under the mentioned transformation groups. 
The statistics $P(S)$ of the level spacings  $S$ changes under transformations with the elements of those 
groups, so it depends on the basis chosen for their representation. Nevertheless, 
they are important in specific physical situations and also involve interesting 
mathematics. 

We shall first treat rigorously the case of general 2D non-normal matrices\footnote{ 
A matrix ${\bf A}$ is non-normal \cite{MehtaMT} if it does not commute with its
adjoint, i. e., $[{\bf A},{\bf A}^+ \underline{]} \neq 0$. Non-normal matrices have
important physical applications, especially in dissipative systems
\cite{GroRob}, \cite{tre93},  \cite{gro00}, \cite{eck06}, \cite{neu97}.}
with Gaussian distributed matrix elements even admitting different variances of the various matrix elements. 
The level repulsion exponent $\rho$ in this case will turn out to be still $\rho = 1$, unless there 
are constraints on the matrix elements 
like e. g. having only positive or only negative nondiagonal elements etc.
More general matrix element distributions in this case are
left for future research, as they introduce additional serious mathematical
problems, cf. section \ref{generalreal}. 

Second we consider {\em symmetric} matrices with other, general
distributions of the real matrix elements. For such symmetric, real matrices we shall show that the level
repulsion exponent is always $\rho = 1$, provided the distributions of
all matrix elements are regular at zero value. Next we study in detail
and without approximations the following specific cases of such regular distributions, 
namely: the uniform (box) 
distribution, the Cauchy-Lorentz distribution, the exponential distribution, 
and also some addenda to the Gaussian case, already dealt with in \cite{GroRob}. 

If, in contrast,
the distribution of matrix elements is singular at zero value,
$P(S)$ shows different behaviour at $S=0$. For example,
if the singularity of the statistics of the matrix elements is an integrable power 
law at zero value of $S$, the level spacing statistics 
$P(S)$ exhibits a fractional exponent pawer law level repulsion as was
discovered and treated in \cite{ProRob1993a},\cite{ProRob1994a},
\cite{ProRob1994b}, \cite{ProRob1993b}. This probably is
characteristic for sparsed matrices, which in turn are important random matrix 
models for nearly integrable systems of the KAM type.

\section{General real 2D matrices}
\label{generalreal}

\subsection{Level spacing statistics}
\label{leveldistancestatistics}

Consider $2 \times 2$ matrices ${\bf A} = (A_{ij})$, where $i,j = 1$ or $2$. 
This matrix has two diagonal elements, which can always be chosen as $a$ 
and $-a$. For general $A_{ij}$ introduce $A_s = \frac{1}{2}\left( A_{11} 
+ A_{22} \right)$ and subtract the diagonal matrix $A_s {\bf \underline{1}}$. 
Then $A_{11} - A_s \equiv a = - (A_{22} - A_s)$, i. e., one obtains the 
formula (\ref{generaltwomatrix}) without loss of generality.
Quite generally, for a matrix $\left( \begin{array}
{cc} a & b\\ c & d \end{array} \right)$ the level spacing 
$S= | \lambda_1-\lambda_2 | = | \sqrt{(a-d)^2 + 4bc} ~|$ only depends on 
the difference $a-d$, so that we can arbitrarily shift $a$ and $d$ 
by a constant, in particular by $A_s$.
\footnote{Let us make clear that in the general symmetric GOE matrix
$\left( \begin{array} {cc} a & b\\ b & d \end{array} \right)$ the 
variances of the diagonal elements $a$ and $d$ are equal, but
by a factor 2 larger than the variance of the offdiagonal element $b$.
However, setting  $d=-a$ implies, that the GOE case occurs when
the variance of $a$ is equal to the variance of $b$. See also
subsection 4.1}

Let $a$ as well as the nondiagonal elements $b_1$ and $b_2$ all be real and write

\be \label{generaltwomatrix}
{\bf A} = \left(A_{ij}\right) = \left(  \begin{array} {cc}
a & b_1  \\ b_2 & -a
\end{array} \right).
\ee
If $b_1 = b_2$, the matrix ${\bf A}$ is symmetric.

The eigenvalues of ${\bf A}$ follow from
\be \label{eigenvalue equation}
\left| {\bf A} - \lambda {\bf \underline{1}} \right| =
\left| \begin{array}{cc}  a - \lambda  &  b_1  \\  b_2  &  -a - \lambda 
\end{array} \right|
= \lambda^2 - a^2 - b_1 b_2 = 0    \ ,
\ee

\be \label{eigenvalues}
\mbox{i. e.,} \ \ \lambda_{1,2} = \pm \sqrt{a^2 + b_1 b_2} \ .
\ee
The eigenvalues are real for arbitrary symmetric ($b_1 = b_2$) matrices.
In the more general case of $b_1 \neq b_2 $ the eigenvalues are still real, 
if the product $b_1 b_2$ is larger than $-a^2$ , otherwise they are
purely imaginary, but never general complex. 

If the matrix ${\bf A}$ is {\bf not} symmetric, it is 
no longer normal. Namely, in the general case one finds for the commutator
\be \label{commutator}
[{\bf A}, {\bf A}^+ \underline{]} = \left( \begin{array} {cc}
b_1^2 - b_2^2 ~ ~ ~& 2a(b_2 - b_1) \\ 2a(b_2 - b_1) ~ ~ ~& 
b_2^2 - b_1^2
\end{array} \right).
\ee
Apparently the commutator $[{\bf A}, {\bf A}^+ \underline{]} = 0$ is zero iff
$b_2 = b_1$, i. e., in the symmetric case. In general, $2 
\times 2$ matrices {\bf A} are non-normal, $[{\bf A}, {\bf A}^+ \underline{]} \neq 0$.

The distribution $P(S)$ of the level spacings  $S$ is given by

\be \label{generalP(S)}
P(S) = \int_{- \infty}^{+ \infty} \int_{- \infty}^{+ \infty} \int_{- 
\infty}^{+ \infty} da ~db_1 ~db_2 ~\delta \left( S - 2 \left| \sqrt{a^2 +  b_1  b_2 } \right| \right) 
g_a(a) g_{b_1}(b_1) g_{b_2}(b_2)   .
\ee
Here $\delta(.)$ is the Dirac  delta function and 
$g_a(a),g_{b_1}(b_1),g_{b_2}(b_2)$ are the normalized probability 
density functions for the matrix elements $a, b_1, b_2$, respectively.
$P(S)$ is the central object of our study in this paper. We are going
to study the dependence of the main properties of $P(S)$, especially
the small-$S$ behaviour (the level repulsion) as well as the
asymptotic behaviour at large $S$ (the tail of $P(S)$), upon the
main features of the matrix element distribution functions 
$g_a(a),g_{b_1}(b_1),g_{b_2}(b_2)$. 
In the special case of an ensemble of random symmetric matrices ${\bf A} = {\bf A}^+$ we must
have $b_1=b_2$. It is not enough that the two statistics $g_{b_1}$ and
$g_{b_2}$ are equal! This is achieved by inserting in the
integrand of (\ref{generalP(S)}) the constraint $g_{b_2}(b_2) = \delta (b_1-b_2)$.
Integrating then over $b_2$ results in the level distribution formula 

\be \label{generalsymmetricP(S)}
P(S) = \int_{- \infty}^{+ \infty} \int_{- \infty}^{+ \infty} 
 da ~db_1 ~\delta \left(S - 2 \sqrt{a^2 +  b_1^2 }\right) 
g_a(a) g_{b_1}(b_1).
\ee

We shall work out exact formulae for the following cases: (i) General, non-normal 
matrices with Gaussian distributed elements $a,b_1,b_2$ in sections 
\ref{generalreal} and \ref{general_non-normal}, and (ii)  
symmetric (normal) matrices but considering non-Gaussian 
distributions of the matrix elements in section \ref{symmetric}: uniform (constant or box) distribution, 
Cauchy-Lorentz distribution, exponential distribution, and singular distribution 
(integrable power law at $S=0$ multiplied by an exponential tail). Here we also detail more 
on the Gaussian case in order to extend our results of reference \cite{GroRob}. In section 
\ref{triangular} we comment on the level distribution of the prototype non-normal matrix, 
the triangular matrix. The final section \ref{discussions} is devoted to a discussion and conclusions.

\subsection{Polar coordinate representation of the level spacing distribution}
\label{polar_formulae}

We notice that the radicands in the arguments of the delta functions in both 
equations (\ref{generalP(S)}) and (\ref{generalsymmetricP(S)}) are homogeneous 
in the moduli of $a$, $b_1$, and $b_2$. Thus it is natural to introduce spherical 
or plane polar coordinates. In the general case, described by equation (\ref{generalP(S)}), we define 

\be \label{sphericalcoordinates}
b_1=r \cos \theta \cos \varphi,\;\;\; b_2=r \cos \theta \sin \varphi,\;\;\;
a=r \sin \theta ,
\ee
where $r \in [0,\infty)$, $-\pi/2 \le \theta \le \pi/2$, and $0\le \varphi \le 2\pi$.
Then we get for the level distance

\be  \label{radicand}
S = 2 ~\left| \sqrt{a^2 + b_1b_2} ~\right| = 2~r ~Q  ~~\mbox{with} ~~ Q(\theta, 
\varphi) = \left| \sqrt{\sin^2 \theta + \half \cos^2 \theta \sin2\varphi} ~\right|.
\ee
The Jacobian of the coordinate transformation is $r^2 \cos \theta$ and therefore 
$da~db_1~db_2 = r^2 dr~ \cos \theta~ d\theta~ d\varphi$. The
$r$-integration can be carried out in favour of $S$ resulting in

\bea \label{GeneralP(S)spherical}
P(S) = \frac{S^2}{8} \int_{-\pi/2}^{\pi/2} \int_{0}^{2\pi} 
\frac{cos \theta ~d\theta ~d\varphi}{Q^3}  \nonumber  \\ 
\hspace{20mm} \times g_a(\frac{S}{2Q} \sin \theta) g_{b_1}(\frac{S}{2Q}\cos\theta \cos\varphi)
g_{b_2}(\frac{S}{2Q}\cos\theta \sin\varphi).
\eea
Now, if the value of the double integral were regular at $S=0$,
the level repulsion would be quadratic, $P(S) \propto S^2$. But $Q$ has zeros
at $\theta = 0$ together with $\varphi = 0, \pi /2, \pi, 3\pi/2, 2\pi$, which 
are scanned in the integration. That leads to singularities of the integrand,
explicit ones ($1/Q^3$) as well as implicit ones (in the $g$-arguments). In particular 
the $Q^{-1}$ in the $g$'s enforces a scan of the full matrix element distribution functions
including their tails, for any nonzero value of $S$. One cannot simply expand the $g$'s in 
terms of $S$ and so find the small $S$ behaviour. Thus the picture of the $S$-dependence is
far from simple. The small $S$ behaviour depends on all details of the matrix element 
distribution functions $g$ including their tails. Therefore in this general case of independent
$a, b_1, b_2$ we shall choose another approach to attack this problem and analyse it at least 
in the case of Gaussian $g$'s in section \ref{general_non-normal}.

In the case of real symmetric 2D matrices the transformation to plane polar coordinates is 
simpler and thus much more useful. Here the $r$-integration does not produce singularities, 
since the analog of $Q$ here is just $1$. Introduce plane polar coordinates into equation 
(\ref{generalsymmetricP(S)}), 

\be \label{polarcoordinates}
a = r \cos \varphi, \;\;\; b_1= r \sin\varphi ,
\ee
where  $r\in [0,\infty)$ and $0 \le \varphi \le 2\pi$. The Jacobian is $r$, i. e., $da~db_1 = rdr~d \varphi$.
The level distance reduces to $2~\sqrt{a^2 + b_1^2} = 2~r$, which is independent of $\varphi$ in contrast 
to the $\theta, \varphi$ dependent factor $Q$ in the general case of equation (\ref{radicand}).
We now can do the $r$-integration immediately and get

\be \label{SymmetricP(S)polar}
P(S) = \frac{S}{4} \int_0^{2\pi} d\varphi~ g_a(\frac{S}{2}\cos\varphi)
~g_{b_1}(\frac{S}{2}\sin\varphi).
\ee
Here, for $S \rightarrow 0$ one {\em can} use the power law expansions of the matrix 
element distribution functions $g_{a,b_1}$ in terms of $S$, independent of the nature of 
the $g$'s, Gaussian or non-Gaussian. In particular, if $g_a(x)$ and $g_{b_1}(x)$ are 
both regular at $x=0$, the integrand at $S=0$ is just a constant and equal to $g_a(0) g_{b_1}(0)$. 
We obtain for small $S$

\be \label{Symmetriclevelrepulsion}
P(S) \approx S \cdot \frac{\pi}{2} g_a(0) g_{b_1}(0).
\ee
Thus in case of regular $g$'s at zero value we always have linear
level repulsion $P(S) \propto S$ for real, symmetric, random matrices, whose amplitude reflects 
the probability density $g_{a,b_1}(0)$ to find the matrix elements $a=0$ and $b_1=0$ in the matrix $A$. 
Higher order corrections in $S$ can be derived by Taylor expanding the $g$'s around
$x=0$. We conclude that regular matrix element distribution functions $g_{a,b_1}$ transform 
into regular level spacing distributions $P(S)$. From  equations (\ref{SymmetricP(S)polar}) and 
(\ref{Symmetriclevelrepulsion}) we also notice that the level
repulsion is {\bf not} linear if $g_a(0)$ and $ g_{b_1}(0)$ do not exist, i. e., if the 
distributions $g_a(x)$ and $g_{b_1}(x)$ are singular at $x=0$, if there is infinite probability 
density for the matrix elements to have values $x=a=0$ or $x=b_1=0$. We shall study this 
important case in section \ref{symmetric}.

\section{General non-normal real 2D matrix ensemble with Gaussian
distributed matrix elements}
\label{general_non-normal}

\subsection{Level spacing distribution $P(S)$: general}
\label{level general}

We start with equation (\ref{generalP(S)}) and observe that the 
dependence of the level distance $S$ in the integrand of $P(S)$ 
on $b_1$ and $b_2$ is only through the product $B=b_1b_2$.  Therefore 
it is natural to introduce hyperbolic coordinates defined as

\be   \label{hyperboliccoordinates}
B = b_1b_2,\;\;\; v = \frac{b_1}{b_2}, ~~\mbox{equivalent to} ~~b_1^2 = Bv, ~~b_2^2 = B/v.  
\ee
Both $B$ and $v$ run over the entire interval $(-\infty, \infty)$, but always have the same
sign, sgn~$B$ = sgn $b_1 ~\cdot$ sgn $b_2$ = sgn~$v$. Positive $B$ or $v$
indicate that both non-diagonal elements are positive or that both are negative.
Negative values of the variables $B$ and $v$, instead, describe the case of non-diagonal elements 
with different sign. The Jacobian determinant is $J = 1/(2 |v|)$, and for the area elements 
we have $db_1~db_2 = dB~dv/(2|v|)$.  

To analyse the integral further we assume Gaussian distributed matrix elements, though 
with possibly different variances $\s^2, \s_1^2$, and $\s_2^2$, for $a,b_1$, and $b_2$, respectively.

\be \label{Gaussianelements}
g_a(a) = \frac{1}{\sigma \sqrt{\pi}} \exp(-\frac{a^2}{\sigma^2}),\;\;\;
g_{b_i}(b_i) = \frac{1}{\sigma_i \sqrt{\pi}} \exp(-\frac{b_i^2}{\sigma_i^2}), ~~~ i=1,2.
\ee
All three distributions are normalized to one.
With this assumption the equation for $P(S)$ obtains the form

\begin{eqnarray} \label{GaussianP(S)}
P(S) = \frac{1}{\sigma\sigma_1\sigma_2\sqrt{\pi}^{3}}
\int_{-\infty}^{\infty} \int_{-\infty}^{\infty} \int_{-\infty}^{\infty} \frac{da~dB~dv}{2|v|} 
\exp \left(- \left( \frac{a^2}{\sigma^2} + \frac{Bv}{\sigma_1^2} 
+ \frac{B}{v\sigma_2^2} \right) \right)  \nonumber \\
\hspace{60mm}  \times \delta \left(S-2 \left| \sqrt{a^2+B} \right| \right) .
\end{eqnarray}
The integrand is even in the variable $a$, we thus can use $\int_{- \infty}^{\infty} da \rightarrow 2 
\int_0^{\infty} da$. Next, the level distance delta function does not depend on $v$ explicitly. But 
since the signs of $v$ and $B$ are coupled, only the first and the third quadrant of the ($B,v$)-plane 
contribute. In both these ($B,v$)-quadrants it is $b_1^2 = Bv = |B| |v| \ge 0$ and $b_2^2 = B/v = 
|B| / |v| \ge 0$, guaranteeing the convergence of the ($B,v$)-integrals. In the first quadrant we 
have positive $B$, while in the third one $-|B|$ is relevant for the level distance. This leads to 

\begin{eqnarray} \label{Gaussian0}
P(S) = \frac{2}{\sigma \sigma_1 \sigma_2 \sqrt{\pi}^3} \int_0^{\infty} da \exp \left( -\frac{a^2}{\sigma^2} \right) 
\int_0^{\infty} dB \int_0^{\infty} \frac{dv}{2v} \exp \left( - \frac{Bv}{\sigma_1^2} - \frac{B}{v \sigma_2^2} \right) 
\nonumber \\     \hspace{34mm} \times \left[ \delta \left( S - 2 \sqrt{a^2 + B} \right) + 
\delta \left( S - 2 \left| \sqrt{a^2 - B} \right| \right) \right].
\end{eqnarray}
Now all variables $a, B, v$ have to be integrated over positive values only. The sum of the 
delta-functions then is independent of the variable $v$ and the $v$-integral can be performed,
cf. \cite{GR} No. 3.478,4. 

\be \label{modifiedBessel}
\int_0^{\infty} \frac{dv}{2v} \exp \left( - \frac{B}{\sigma_1^2} v - \frac{B}{\sigma_2^2} \frac{1}{v} \right)
= K_0 \left( \frac{2 B}{\sigma_1 \sigma_2} \right).
\ee
Here $K_0(x)$ is the modified Bessel function of second kind and zero order. Its argument is always positive,
$B \ge 0$. The level spacing distribution $P(S)$ becomes 

\be \label{Gaussian1}
P(S) = \frac{2}{\sigma\sigma_1\sigma_2\sqrt{\pi}^{3}} \int_0^{\infty}
dB~ K_0\left( \frac{2 B}{\sigma_1\sigma_2}\right) \cdot G(B),
\ee
where  $G(B)$ is the Gaussian averaged level distribution for fixed, given product
$B \ge 0$ of the nondiagonal elements but varying diagonal elements $a$,

\begin{eqnarray} \label{Gaussian2}
G(B) = \int_{0}^{\infty} da ~\exp \left(- \frac{a^2}{\sigma^2}\right) \nonumber \\
\hspace{15mm} \times \left[ \delta \left(S-2\sqrt{a^2+B} \right) + 
\delta \left(S-2 \left| \sqrt{a^2 - B} \right| \right) \right] , ~ \mbox{with} B \ge 0. 
\end{eqnarray}
The calculation of $G(B)$ is easy because of the delta function and can be done analytically. 
Consider the first delta-function. Its argument, denoted as $f(a) = S - 2 \sqrt{a^2 + B}$, with 
positive $B$, corresponds to real eigenvalues of the matrix {\bf A}. The zeros $a_i$ of the 
delta function contribute to the integral only if they are real and positive. There is only one real,
positive $a_i$, provided $B \le S^2 /4$. 
Then the variable $u = 4B / S^2$ fulfils $0 \le u \le 1$ and the $f$-zero reads 
$a_i= \frac{S}{2} \sqrt{1 - u}$. The weight of the delta function contribution is given by the inverse of the 
derivative of $f$, which is $|f'(a_i)|^{-1} = (2 \sqrt{1 - u} ~)^{-1}$. 
The first delta-function in $G(B)$ then leads to $G_{+re}(B) = \exp (- \frac{S^2}{4 \sigma^2} (1-u)) / 
(2 \sqrt{1-u}~)$. The label ($+re$) indicates that the term describes the case $+B$ and real ($re$) eigenvalues. 
Transforming the variable $B \rightarrow u$ this $G_{+re}(B)$ contributes the following 
term to the level level spacing distribution 

\be \label{Gaussian5}
P_{+re} (S) = \frac{S^2}{4\s\s_1\s_2\sqrt{\pi}^3}
\int_0^1 \frac{du} {\sqrt{1-u}}~ K_0\left( \frac{S^2u}{2\s_1\s_2}\right)
~\exp \left( -\frac{S^2}{4\s^2}(1-u)\right).
\ee
This integral has been considered already in ref.\cite{GroRob} and leads to a level repulsion exponent 
$\rho = 2 - 0_{log}$, see also section \ref{level details}. It will turn out that this contribution 
(\ref{Gaussian5}) is subdominant relative to the other two integrals for $P(S)$, 
which in contrast to (\ref{Gaussian5}) will lead to the repulsion 
exponent $\rho = 1$, cf. again section \ref{level details}.
  
We now calculate the $G(B)$-contributions coming from the second delta function. They are labelled 
by a ($-$) sign (since $-B$ enters) and correspond to real as well as to imaginary eigenvalues of the matrix {\bf A},
depending on the size of $B$. They are labelled therefore by ($re$) if $a^2 - B \ge 0$ and by 
($im$) if $a^2 - B < 0$. The relevant (positive) zeros $a_i$ of the argument of the delta function 
are $a_i = \frac{S}{2} \sqrt{1 + u}$ for all positive $B$ or $u = 4B/S^2 \ge 0$ in case of ($re$) and
$a_i = \frac{S}{2} \sqrt{u - 1}$ for all positive $B$ with $B \ge S^2 / 4$ and thus all 
$u = 4B / S^2 \ge 1$ in the case ($im$). The weights $|f'(a_i)|^{-1}$ of the delta function 
contributions are obtained from $|f'(a_i)| = 2 \sqrt{1 + u}$ for the case ($re$) 
and from $|f'(a_i)| = 2 \sqrt{u - 1}$ for ($im$). These formulae lead to the following two 
contributions to the level spacing distribution $P(S)$:

\be  \label{Gaussian6}
P_{-re} (S) = \frac{S^2}{4\s\s_1\s_2\sqrt{\pi}^3}
\int_0^{\infty} \frac{du} {\sqrt{1+ u}}
~ K_0\left( \frac{S^2 u}{2\s_1\s_2}\right)
~\exp \left( -\frac{S^2}{4\s^2}(1+ u)\right)
\ee
and

\be  \label{Gaussian7}
P_{-im} (S) = \frac{S^2}{4\s\s_1\s_2\sqrt{\pi}^3}
\int_1^{\infty} \frac{du} {\sqrt{u-1}}
~ K_0\left( \frac{S^2u}{2\s_1\s_2}\right)
~\exp \left( -\frac{S^2}{4\s^2}(u-1)\right).
\ee

These three integrals (\ref{Gaussian5}), (\ref{Gaussian6}), and (\ref{Gaussian7}) can be 
summed up to give the complete level spacing distrribution $P(S)$. To do this 
one introduces variables $y$ such that in all three cases the exponential is $\exp (-y)$ 
with $y \ge 0$. In the case ($-im$) one in addition substitutes $y \rightarrow -y$. 
The real eigenvalues, represented by equations (\ref{Gaussian5}) and (\ref{Gaussian6}), 
give $y$-integrations from 0 to $S^2 / 4 \sigma^2$ and from $S^2 / 4 \sigma^2 $ to $\infty$. 
The imaginary eigenvalues lead to an integral from $0$ to $\infty$ or, equivalently, from
$- \infty$ to 0. 
Respecting the always positive argument of $K_0$ one evaluates for the complete level spacing distribution function in closed form

\be \label{Gaussian8}
P(S) = \frac{S}{2 \s_1\s_2\sqrt{\pi}^3}
\int_{-\infty}^{\infty} \frac{dy} {\sqrt{|y|}}
~ K_0\left( \frac{2\s^2}{\s_1\s_2} \left| y - \frac{S^2}{4\s^2} \right| \right)
~\exp ( -|y|) .
\ee

This formula has been reported to us independently by Professor H.-J. Sommers
\cite{Somm}. Because the integral converges also for $S = 0$, we can conclude from this 
formula that in leading order in $S$ we have linear level repulsion due to the explicit 
factor $S$, namely 

\be \label{Gaussian9}
P(S) =  S \cdot \frac{1}{\s_1\s_2\sqrt{\pi}^3}
\int_{0}^{\infty} \frac{dy} {\sqrt{y}}
~ K_0\left( \frac{2\s^2}{\s_1\s_2} y \right)
~\exp ( -y) ~ + ~\mbox{h.o.t.}
\ee
The $S$-independent integral can be expressed in terms
of the hypergeometric function $F$ as will be shown in equation (\ref{2F1}) of section \ref{level details}.

From equation (\ref{Gaussian8}) one might wish to work out also the higher order terms in $S$,
stemming from the integrand.
However, if at small $S$ one formally expands the factor $K_0$ in (\ref{Gaussian8})
in terms of an $S^2$-series, one obtains contributions to the integrand, which are no longer
integrable. Therefore the small $S$ behaviour of the $y$- or $u$-integrals is highly nontrivial.
This agrees with our earlier observation in section \ref{polar_formulae}.
We have to analyse that in detail by studying the individual
integrals $P_{+re}(S)$, $P_{-re}(S)$, and $P_{-im}(S)$
of equations (\ref{Gaussian5}), (\ref{Gaussian6}), and (\ref{Gaussian7}), respectively.

\subsection{Level spacing distribution $P(S)$: Details of small $S$ behaviour}
\label{level details}

As we have seen in sections \ref{polar_formulae} and \ref{level general} for non-symmetric 
(and thus non-normal) matrices, the behaviour of $P(S)$ at small $S$ is very delicate and
depends on the details of the distribution functions $g_{a,b_i}(x)$ for the matrix elements.
In order to achieve understanding we go back to equation 
(\ref{Gaussian5}) and make the substitutions $1-u \rightarrow u' \rightarrow u$ and 
$S^2u/(2\s_1\s_2) = u' \rightarrow u$ yielding

\be \label{real1}
P_{+re} (S) = \frac{S}{2 \s \sqrt{2\s_1\s_2} \sqrt{\pi}^3}
\int_0^{\ep}  \frac{du}{\sqrt{u}} ~K_0(\ep-u) ~e^{-Au}.
\ee
Here we have introduced the notations

\be \label{defepsilonB}
\ep = \frac{S^2}{2 \s_1\s_2},\;\;\; A = \frac{\s_1\s_2}{2\s^2}.
\ee
If $\ep$ is very small, $\ep \ll 1$, we can use the approximation
of $K_0(z)$ at small $z$, which is (\cite{GR}, No. 8.447,1 and 3,
and 8.362,3), $K_0(z) = -\ln\frac{z}{2}- C + O(z)$, where
$C$ is Euler's constant $0.577215\dots$. 
Then the leading term in the above integral, also
taking into account the Taylor expansion of $e^{-Au}$,  can be written as

\be \label{real1.1}
\hat{I}_{+re}(\ep) = \int_0^{\ep} \frac{du}{\sqrt{u}}  
\left( -\ln(\ep-u)\right), 
\ee
which after a simple substitution $\ep - u = u' \rightarrow  u$
can be found in \cite{GR} (No. 2.727,5). After
the evaluation for small $\ep$ we get

\be \label{real1.2}
\hat{I}_{+re}(\ep) \approx 2\sqrt{\ep} (- \ln \ep ) \propto S\ln S^{-2},
\ee
meaning that including the explicit factor $S$ in (\ref{real1}) we have 

\be \label{real_repulsion}
P_{+re} (S) \propto S^2 \ln S^{-2}.
\ee
The level repulsion exponent $\rho$ for the case of positive $B$ (which also guarantees real 
eigenvalues of the matrix {\bf A}) therefore is $\rho = 2-0_{log}$. Positive $B$ means that we consider 
Gaussian distributed non-normal, real 2D matrices, which have only positive or only negative non-diagonal matrix 
elements $b_1, b_2$. It is under this constraint that they have the rather strong repulsion exponent 
$\rho = 2 - 0_{log}$, as reported already in reference \cite{GroRob}. 

The other two contributions (\ref{Gaussian6}) and (\ref{Gaussian7})
are different in behaviour. Indeed each of them gives rise
to weaker level repulsion, i. e., to a smaller exponent $\rho$. They both exhibit linear level
repulsion $P(S) \propto S$, thus $\rho = 1$. This dominates the stronger, quadratic repulsion 
for small $S$ valid for $P_{+re}(S)$, 
so that the total $P(S)$ has linear level repulsion $P(S) \propto S$,
as obtained in (\ref{Gaussian8}) and (\ref{Gaussian9}). The analytical reason for the quite different 
$S$-dependence of $P_{+re}(S)$, from equation (\ref{Gaussian5}), in contrast to that of 
$P_{-re}$ and $P_{-im}$, according to equations (\ref{Gaussian6}) and 
(\ref{Gaussian7}), is the {\bf finite} range of the $B$- or $u$-integration in (\ref{Gaussian5}) versus the 
infinite integration intervals in the cases of negative $B$, 
namely (\ref{Gaussian6}) and (\ref{Gaussian7}). Because of these infinite  
integration intervals one cannot Taylor expand the $K_0$-function in the cases of 
$P_{-re}$ and $P_{-im}$. Instead, its complete functional form including its 
tails affect the convergence and thus the $S$-behaviour of the integrals 
in (\ref{Gaussian6}), (\ref{Gaussian7}).

In order to demonstrate this explicitly, we use the same substitution as before
and arrive from equation (\ref{Gaussian6}) at the following expression

\be \label{real2}
P_{-re} (S) = \frac{S \exp( -\frac{S^2}{4\s^2})}
{2 \s \sqrt{2\s_1\s_2} \sqrt{\pi}^3}
\int_0^{\infty}  \frac{du}{\sqrt{u + \ep}} K_0(u) e^{-Au},
\ee
with the relevant integral

\be \label{real2.1}
\hat{I}_{-re} (S) = 
\int_0^{\infty}  \frac{du}{\sqrt{u + \ep}} K_0(u) e^{-Au}.
\ee
The meaning of $\ep$ and $A$ is the same as in equation (\ref{defepsilonB}). The integrand decays 
exponentially for large $u$ since also $K_0(u)$ does so. Thus the upper limit of
the integral converges safely, independent of $\ep$. For small
$u$ the integrand can be estimated in a similar way as before. The
result is that the integral converges to a {\em finite} value for $\ep \rightarrow 0$ and not to zero,
as in equations (\ref{real1.1}) and (\ref{real1.2}) for $\hat{I}_{+re}(S)$, i. e.,  
$\hat{I}_{-re} (S \rightarrow 0) \ne 0$ is finite as $S\rightarrow 0$. Consequently,
from the explicit factor $S$ in equation (\ref{real2}), we conclude that the level 
repulsion is linear, $\rho = 1$. Note that in contrast
to the finite integration range in the integral (\ref{real1.1}), which for small $S$ behaves 
$\propto S \log S^{-2}$, the infinite range integral (\ref{real2.1}) 
has a {\em finite}, nonzero limit for $S \rightarrow 0$.  

It remains to estimate the contribution from the imaginary eigenvalues 
given by (\ref{Gaussian7}). Again, similar substitutions of the integration 
variable $u$ lead us to the expression

\be \label{imag}
P_{-im} (S) = \frac{S}{2 \s \sqrt{2\s_1\s_2} \sqrt{\pi}^3}
\int_0^{\infty}  \frac{du}{\sqrt{u}} K_0(\ep+u) e^{-Au}.
\ee
$\ep$ and $A$ are defined in equations (\ref{defepsilonB}). As before, the integral has a finite 
value, which does not go to zero with $\ep \propto S^2 \rightarrow 0$, and thus 
the level repulsion of the contribution $P_{-im}$ according to equations  (\ref{Gaussian7}) and (\ref{imag}) 
is again linear, $\rho = 1$. In fact, the integral in (\ref{imag})
at $\ep=0$ can be calculated according to \cite{GR} (6.621,3)
as

\be \label{2F1}
\int_0^{\infty}  \frac{du}{\sqrt{u}} K_0(u) e^{-Au} = 
\frac{\sqrt{\pi}^3}{\sqrt{1+A}} ~
F \left( \half, \half; 1; \frac{A-1}{A+1} \right) \ ,
\ee
where $F$ is the hypergeometric function.

The conclusion is that the level repulsion of the complete level spacing
distribution  function $P(S)$ is linear, $\rho = 1$,
due to the contributions $P_{-re}(S)$ and $P_{-im}(S)$, 
whereas the contribution $P_{+re}(S)$ alone has the level repulsion
exponent $\rho = 2-0_{log}$, as we have already found in \cite{GroRob} for the case of Gaussian
random matrices with only positive or only negative nondiagonal elements. 
Note that the criterion for the different small $S$ 
behaviour is, if the product $b_1b_2 = B$ is always positive, $B \ge 0$, or if there are
also $B < 0$ contributions. The eigenvalues according to this latter case are apparently more abundant
and thus have less repulsion ($\propto S$), while the former ones with only a positive product 
$b_1b_2$ lead to the repulsion behaviour $\propto S^2 \log S^{-2}$.

\section{Symmetric (normal) real 2D matrix ensembles with various 
distributions of the matrix elements}
\label{symmetric}

In this section we treat 2D real random matrices which are {\em symmetric}, i. e., 
$b_1 = b_2 \equiv b$ and thus $B = b^2 \ge 0$ always (from here onwards we drop 
the labels 1 and 2 in $b_1 = b_2$). Such matrices are normal, i. e., the matrix 
commutates with its adjoint. The generalization here is, that we allow for 
a broad variety of matrix element distribution functions $g_{a,b}$. 
The following classes of matrix element distributions are considered: (1) Gaussian distribution revisited,
(2) box (uniform) distribution, (3) Cauchy-Lorentz distribution, (4) exponential
distribution, and (5) singular distribution (power law approaching zero
value, multiplied by exponential tail). In all these cases we shall 
start with equation (\ref{SymmetricP(S)polar}) as a useful integral representation for
$P(S)$ in terms of $g_{a,b}$. 

\subsection{Gaussian distribution revisited}
\label{gaussian distribution}

This case of real, Gaussian distributed 2D matrices with possibly different widths of the 
diagonal and the non-diagonal matrix element statistics has been treated recently 
in reference \cite{GroRob}. We briefly
repeat it here for the sake of completeness. Using the normalized Gaussians 
defined in (\ref{Gaussianelements}) we immediately get 

\be \label{GaussianP}
P(S) = \frac{S}{2\s_a\s_b} e^{-\frac{S^2}{8} (\s_a^{-2}+\s_b^{-2})}
I_0 \left(\frac{S^2}{8} (\s_a^{-2}-\s_b^{-2})\right),
\ee
where $I_0(z)$ is the modified Bessel function of the first kind and zero
order. According to \cite{GR}(No. 8.447,1) 
its small-z expansion is 
$ I_0(z) = 1 + \frac{z^2}{4} + \frac{z^4}{64} + O(z^6)$.
Thus $I_0(0) = 1$ and the level repulsion is linear, $\rho = 1$.
The details for the  general case of different widths of the $a$ and $b$ statistics, 
$\s_a \not= \s_b$, has been analyzed and discussed in \cite{GroRob}. 
We mention that in the case of equal statistics of all matrix elements $\s_a = \s_b = \s$ 
we get, of course, the well known 2D GOE result 

\be \label{Wigner}
P(S) =  \frac{S}{2\s^2} e^{-\frac{S^2}{4\s^2}}.
\ee
After normalizing the first moment to unity, $\langle S \rangle = 1$,
leading to $\s = 1/\sqrt{\pi}$, the level spacing distribution $P(S)$ becomes 
the Wigner distribution $P_{Wigner}(S) =\frac{\pi S}{2} \exp (- \pi S^2/4)$.

\subsection{Box (uniform) distribution}
\label{box distribution}

Let us now consider very different matrix element distributions $g_{a,b}$. We start by studying 
the following uniform distributions for the matrix elements $a$ and $b$,

\be \label{boxdistributionsa}
 g_a(a) = \frac{1}{2a_0}, \;\;\;  \mbox{if $|a| \le a_0$},
\;\;\; 0\;\;\; \mbox{otherwise} \ ,
\ee
\be \label{boxdistributionsb}
 g_b(b) = \frac{1}{2b_0}, \;\;\;  \mbox{if $|b| \le b_0$},
\;\;\; 0\;\;\; \mbox{otherwise} \ .
\ee
Thus the probability density product $g_a(a)g_b(b)$ is constant, equal to
$(4a_0b_0)^{-1}$, inside the
(centrally located) rectangle with sides $2a_0 \times 2b_0$, and therefore
for $S$ smaller than $2 \min\{a_0,b_0\}$ the level spacing distribution $P(S)$ 
can be calculated exactly. It is equal to

\be \label{boxP}
P(S) = \frac{\pi S}{8a_0b_0}, \;\;\;\mbox{if $S \le 2 \min\{a_0,b_0\}$}.
\ee
Again the level repulsion is linear. This is consistent with our finding in section 
\ref{polar_formulae} that whenever $g_{a,b}(0) = const \neq 0$ there is generically linear level 
repulsion $\rho = 1$. From this geometrical picture it is also clear that $P(S)$ is zero for 
$S \ge 2 \sqrt{a_0^2+b_0^2}$. For $S$ in between $P(S)$ varies continuously. 

Indeed, we can calculate
$P(S)$ exactly for all $S$, using equation  (\ref{SymmetricP(S)polar}). 
We observe that $g_a$ and $g_b$ enter this expression symmetrically.
Therefore without loss of generality we assume that $b_0  \le a_0$, i.e.
$\min\{a_0,b_0\} =b_0$. Then we have to consider two more intervals, namely
(i) $2b_0 \le S \le 2a_0$  and  (ii) $2 a_0 \le S \le 2 \sqrt{a_0^2+b_0^2}$.

In the first case (i) the circle of radius $S/2$ intersects the rectangle
at four points. The angle enclosed by the polar ray and the
abscissa is equal to $\varphi_0$, where  $ S \sin \varphi_0 = 2 b_0$.
Therefore, the total length of the $\varphi$-interval 
contributing to the integral in (\ref{SymmetricP(S)polar}) 
is just $4 \varphi_0$,
and consequently  $P(S) =  \frac{S}{4a_0b_0} \arcsin \frac{2 b_0}{S}$.

In the second case (ii) the circle of radius $S/2$ intersects the rectangle
at eight points, each pair of intersection points defining one $\varphi$-interval
where we get the contribution to the integral. But all four angles are
the same due to the double reflection symmetry of the rectangle and circle.
The larger angle $\varphi_2$ between the polar ray and the abscissa is geometrically determined 
by  $S \sin \varphi_2= 2 b_0$, and the smaller one by $S \cos \varphi_1 = 2 a_0$.
Thus for each pair the length of the $\varphi$-interval is equal to $\varphi_2-\varphi_1$,
and since there are four such intervals, the total length of the interval
contributing to the integral is  $4 (\varphi_2-\varphi_1)$.

Putting all together we obtain the following exact result for the level 
spacing distribution $P(S)$
in the case of the uniform (box) distributions  $g_{a,b}$,

\bea  \label{boxP(S)}
P(S) = \left\{ \begin{array}{ll}
      \frac{\pi S}{8a_0b_0},  & \mbox{if $S\le 2 b_0 \le 2 a_0$} , \\
      \frac{S}{4a_0b_0} \arcsin \frac{2b_0}{S}, & \mbox{if $2b_0 \le S\le  2 a_0$} , \\
      \frac{S}{4a_0b_0} \left( \arcsin \frac{2b_0}{S} - \arccos \frac{2a_0}{S}\right),
              & \mbox{if $2a_0 \le S\le 2 \sqrt{a_0^2 +b_0^2}$} , \\
       0, & \mbox{if $S \ge 2 \sqrt{a_0^2 +b_0^2}$} .
       \end{array}
    \right. 
\eea 
If instead of $b_0 \le a_0$ one has $b_0 \ge a_0$, simply interchange $a_0$ and $b_0$ in the above formulae.

\subsection{Cauchy-Lorentz distribution}
\label{cauchy-lorentz}

The normalized probability densities for the matrix elements are defined by

\be \label{Cauchydistr}
g_a(a) = \frac{1}{\pi a_0 ( 1 + \frac{a^2}{a_0^2})},\;\;\;
g_a(b) = \frac{1}{\pi b_0 ( 1 + \frac{b^2}{b_0^2})} \ .
\ee
From equation (\ref{SymmetricP(S)polar}) and using plane polar coordinates we obtain

\be \label{Cauchyint}
P(S) = \frac{S}{4\pi^2 a_0b_0} \int_0^{2\pi} 
\frac{d\varphi}{ (1 + \frac{S^2}{4a_0^2} \cos^2\varphi) 
(1 + \frac{S^2}{4b_0^2} \sin^2\varphi)  } \ .
\ee
The integral for $S \rightarrow 0$ gives $2 \pi$, so that at small $S$ 
we have $P(S) \approx S/(2\pi a_0 b_0)$ in accordance
with equation (\ref{Symmetriclevelrepulsion}). But the integral 
(\ref{Cauchyint}) can also be done exactly. The result is

\be \label{CauchyP}
P(S) = \frac{S}{2\pi a_0b_0} \cdot 
\frac{  \als \sqrt{1+\bets} + \bets\sqrt{1+\als} }
{(\als + \bets +\als \bets) \sqrt{1+\als}\sqrt{1+\bets}} \ ,
\ee
where $\als =S^2/(4a_0^2)$ and $\bets= S^2/(4b_0^2)$. The asymptotic
behaviour of $P(S)$ at large $S$, i. e., $\als \gg 1$ {\em and} $\bets \gg 1$,
is an inverse quadratic power law,

\be \label{Cauchyasympt}
P(S) \approx \frac{4 (a_0+b_0)}{\pi S^2}.
\ee
If $a_0 = b_0 = a$, the complete formula for all level distances $S$ reads

\be \label{Cauchycomplete}
P(S) = \frac{S}{2 \pi a^2} \cdot \frac{1}{(1+\frac{\als}{2}) \sqrt{1 + \als}}, ~\mbox{with} 
~\alpha^2 = S^2 / (4 a^2) .
\ee 
This expression for the level spacing statistics evidently mirror images the Cauchy-Lorentz 
distribution of the ($g_b = g_a$)-statistics in the $P(S)$-statistics.

It is interesting to note that $P(S)$ in (\ref{CauchyP}) 
has a divergent (infinite) first moment, as is clearly seen from the 
asymptotics (\ref{Cauchyasympt}). 
A generalized power law statistics of the type $g_a(a) = C_a/(1 + (a/a_0)^q)$,
with $q=4,6,\dots$, however, has a finite first moment $\langle S \rangle < \infty$. It
will be treated elsewhere.

\subsection{Singular times exponential distribution}
\label{singularandexponential}

Our normalized distributions in this subsection are chosen as 

\be \label{singular}
g_a(a) = C_a |a|^{-\mu_a} e^{-\la_a |a|},\;\;\;
g_b(b) = C_b |b|^{-\mu_b} e^{-\la_b |b|},
\ee
where the normalization constants are

\be \label{constantsC}
C_i = \la_i^{1-\mu_i}/(2 \Gamma(1-\mu_i)). 
\ee
Here $i=a,b$, the exponents $\mu_i < 1$, and $\Gamma(x)$ is the gamma function.
These distribution functions are singular but integrable power laws for
$a,b \rightarrow 0$ and decay nearly exponentially in the tails. Using 
equation (\ref{SymmetricP(S)polar}) and the reflection symmetry (evenness) 
of both distributions $g_{a,b}$ in (\ref{singular}) we get (note that $S$ is positive only, $S \ge 0$)

\be \label{singularP}
P(S) =C_a C_b ~S \left(\frac{S}{2}\right)^{-(\mu_a+\mu_b)}
\int_0^{\pi/2} \frac{ d\varphi~ 
\exp\left( -\frac{S}{2}( \la_a\cos\varphi + \la_b\sin\varphi ) \right) }
{\cos^{\mu_a}\varphi ~\sin^{\mu_b}\varphi }.
\ee
We did not succeed to calculate this integral analytically in closed form.
However, one can evaluate it for small argument $S$, i. e., $S \rightarrow 0$, where the exponential can 
be approximated by $1$ (equivalent to $\lambda_a, \lambda_b \rightarrow 0$, i. e., no 
tail effects in this small $S$ range). The integral then is

\be \label{smallSintegral}
\int_0^{\pi/2} \frac{ d\varphi~ } {\cos^{\mu_a}\varphi ~\sin^{\mu_b}\varphi } =
\frac{ \Gamma (\half - \mu_a)\Gamma(\half - \mu_b)} {2 \Gamma(1- \frac{\mu_a}{2} - \frac{\mu_b}{2} )} \ .
\ee
From this we get the following level repulsion law, now being a fractional exponent power law,

\be  \label{repelsingular2}
P(S) =C_a C_b ~S \left(\frac{S}{2}\right)^{-(\mu_a+\mu_b)}
 \frac{ \Gamma (\half - \mu_a)\Gamma(\half - \mu_b)}
{2 \Gamma(1- \frac{\mu_a}{2} - \frac{\mu_b}{2} )}.
\ee
The power law distribution for the matrix elements leads to a corresponding power law
for the level spacing distribution. The power law exponents $\mu_a$ and $\mu_b$ of the 
matrix element distribution functions $g_{a,b}$ immediately transform 
into the level repulsion exponent. More precisely, the level repulsion exponent is $\rho = 1 - \mu_a - \mu_b$.  
We emphasize that $P(S)$ at $S=0$ is integrable, if the matrix element distributions $g_a(a)$ and $g_b(b)$ 
are integrable at $a=0$ and $b=0$, respectively. 

The physical interpretation of this repulsion exponent $\rho = 1 -  
\mu_a - \mu_b$ comprises two different possibilities, depending on the  
size of the singularity exponents $\mu_a$ and $\mu_b$ of the  
distribution functions $g_a , g_b$ for the matrix elements. If the  
singularities are strong, more precisely, if $\mu_a + \mu_b > 1$, the  
repulsion exponent $\rho$ is negative. This means that due to the  
rather strong sparsing of the matrix there is not a repulsion but,  
instead, an enhancement of the level distance $S = 0$, the zero  
eigenvalues are emphasized. If the singularities are weak, $\mu_a +  
\mu_b < 1$, the phenomenon of level repulsion remains, $\rho > 0$,  
despite the singularities for the matrix element distributions at  
$a=0$ and $b = 0$.

The diagonal elements $\pm a$ and the non-diagonal elements $b$  
determine the level distance with equal weight since $S = 2 \sqrt{a^2  
+ b^2}$. It is for this reason that it is just the sum of the  
singularity exponents which determines $\rho$, giving equal weight to  
the singularities of the diagonal and non-diagonal elements to the  
repulsion. There are two typical limiting cases. (i) $\mu_a = \mu_b  
\equiv \mu$, both singularities are of equal strenght, or (ii) $\mu_a  
\equiv \mu_0 \neq 0$ and $\mu_b =0$ or vice versa. Only one of the two  
matrix element distributions is singular while the other one remains  
regular. Then only the non-diagonal is sparsed and the diagonal is  
regular or the other way round. In the first case (i), if $\mu < 1/2$  
we have still level repulsion, $\rho > 0$, while for $\mu > 1/2 $ (but  
still less than 1) we find enhancement at zero distance $S = 0$. -- In  
the second case (ii) there will always be level repulsion despite the  
singularity at zero for either the non-diagonal or the diagonal  
distribution, since $\rho = 1 - \mu_0 > 0$ always. The repulsion  
exponent in this case will be between 0 and 1. Sparsing like this is one of  
the possible causes that the repulsion exponent scans the interval $(0,1)$ .

These results are interesting in the context
of quantum chaos of nearly integrable (KAM type) systems. As has
been observed in a variety of different systems of mixed type,
at small energies one finds the so-called fractional exponent power 
law level repulsion, well described by the Brody distribution 
rather than by the Berry-Robnik distribution \cite{BR}-\cite{MalPro2002}, 
which in turn has been clearly demonstrated to apply at sufficiently large
energies, i. e., at sufficiently small effective Planck constant. 
The observed deviation from the Berry-Robnik behaviour is due to the
localization and tunneling effects, and is a subject of intense 
current research \cite{GroRobStoVid}. Phenomenologically it
has been discovered in \cite{ProRob1993a}, further developed in
\cite{ProRob1994a}, \cite{ProRob1994b}, and the connection 
with sparsed matrices was established in \cite{ProRob1993b}.
Quite generally, the matrix representation of a Hamilton operator
of a nearly integrable (KAM type) system in the basis of the
integrable part results in a sparsed banded matrix with
nonzero diagonal elements \cite{ProRob1993b}. Such a sparsed matrix
is precisely characterised by the fact that many matrix elements are zero.
In other words, the probability distribution function of the nondiagonal
matrix elements $g_b(b)$  is singular at zero value of $b$, 
in a manner described by equation (\ref{singular}), while the diagonal
matrix elements have a regular distribution function,
which is precisely the case (ii) discussed above.
Our 2D random matrix theory with such singular
matrix element distribution functions therefore predicts qualitatively
a fractional exponent power law level repulsion, the phenomenon observed 
in the above mentioned works \cite{ProRob1993a}-\cite{ProRob1993b}.
Thus we see that the study of random matrices with other than the invariant
ensembles (GOE and GUE) is very important and connects to new
physics. We leave this direction of research for further studies 
in the near future.

The large $S$ behaviour of $P(S)$ in this case is obviously dominated
by the exponentials. Although an exact solution of the integral (\ref{singularP})
is not known, we shall show by applying the mean value theorem to the integral 
that $P(S)$ decays roughly exponentially at large $S \gg 1$.
This will be analysed in the next subsection, devoted to pure exponential matrix
element distributions without power law singularities at the origin.

\subsection{Exponential distribution}
\label{exponential_distribution}

Here we start with the distribution functions of the previous subsection,
but without singularities. I. e., we assume $\mu_a = \mu_b =0$, yielding the case 
of purely exponential distribution of matrix elements. 

\be \label{exponential}
g_a(a) = C_a e^{-\la_a |a|},\;\;\;
g_b(b) = C_b e^{-\la_b |b|}, \;\;\; \mbox{with} ~~ C_i = \la_i/2.
\ee
The level spacing distribution function in this case is

\be \label{exponentialP}
P(S) =C_a C_b ~S 
\int_0^{\pi/2} d\varphi~ 
\exp\left( -\frac{S}{2}( \la_a\cos\varphi + \la_b\sin\varphi ) \right).
\ee
We could not evaluate this analytically in closed form. For small $S$ the linear 
level repulsion law with $\rho = 1$ is recovered, of course,

\be \label{exponential1}
P(S) \approx \frac{\pi C_aC_b}{2} ~ S  = \frac{\pi \la_a\la_b}{8} ~ S.
\ee
At large $S$ we can estimate the integral (\ref{exponentialP}) using the mean 
value theorem. First we write 
$ \la_a\cos\varphi+\la_b\sin\varphi = \hat{A} \sin(\varphi + \phi$), with
$\phi = \arctan (\la_a/\la_b)$, and $\hat{A} = \sqrt{\la_a^2 +\la_b^2}$.
We then substitute the integration variable from $\varphi$ to $\chi = \varphi +\phi$,
which now runs from $\phi$ to $\phi + \pi/2$. This transforms $P(S)$ into

\be \label{exponential2}
P(S) = C_aC_b ~S \int_{\phi}^{\phi + \pi/2} d \chi ~ e^{-\frac{S}{2} \hat{A} \sin \chi}.
\ee
Since the integrand is continuous and
bounded we can apply the mean value theorem, saying that there is
a value $\chi_0 (S)$ in the interval between $\phi$ and $\phi + \pi/2$ such that
the level spacing distribution is

\be \label{meanvalue}
P(S) =  \frac{\pi C_aC_b}{2} \cdot S \cdot e^{-\frac{S}{2} \hat{A} \sin \chi_0 (S)} .
\ee
$\chi_0 (S)$ is expected to be only weakly dependent on $S$.
In this sense the tail of $P(S)$ is roughly exponential.

\section{Triangular matrix}
\label{triangular}

A special, prototype case of a non-normal matrix {\bf A} from equation (\ref{generaltwomatrix}) is the triangular matrix

\be \label{tridiagonalmatrix}
{\bf A} = \left(A_{ij}\right) = \left(  \begin{array} {cc}
a & b  \\ 0 & -a
\end{array} \right).
\ee
Such matrices have been studied in detail in reference \cite{GroRob}. For them $b_1b_2 = B = 0$ 
and the level distance $S$ does not depend on $b$. With the constraint $g_{b_2}(b_2) = \delta (b_2)$ 
in equation (\ref{generalP(S)}) the $a$ and ($b_1 = b$)-integrations factorize, 
leading to the level distribution function

\be \label{triangularP(S)1}
P(S) = \int_{- \infty}^{\infty} da ~ \delta (S- 2|a|) g_a(a) = \frac{1}{2} \left[ g_a \left(\frac{S}{2} \right) 
+ g_a \left( -\frac{S}{2} \right) \right].
\ee
For even distributions $g_a(a)$ we have

\be \label{triangularP(S)2}
P(S) = g_a \left( \frac{S}{2} \right), ~~~~ S \ge 0, ~~~~\mbox{normalized on} ~~[0 \le S \le \infty) \ .
\ee
Apparently for triangular matrices the level spacing distribution is the immediate 
mirror of the diagonal element distribution $g_a(a)$. If $g_a(a)$ is Gaussian, there 
is no level repulsion at all, the repulsion exponent is $\rho = 0$ (see already \cite{GroRob}), though 
still the $S$ are Gaussian and not Poissonian distributed as in the integrable case. 
If $g_a(a)$ is Poissonian, so is $P(S)$. And if 
$g_a(a)$ is centered about some finite value $\bar{a}$, this also holds for the level spacing
distribution. If the $a$-distribution does not include the origin, there even is formally {\em infinite
level repulsion}. For general matrices with non-zero mean values $\bar{a}$ and $\bar{b}_{1,2}$ the level spacing 
distribution is somewhat more tricky.

\section{Discussion and conclusions}
\label{discussions}

We have studied the level spacing distribution of general 2D real 
random matrices. First general, non-normal random matrices with
Gaussian distribution of the matrix elements have been considered, 
showing that in general the level repulsion is still linear, as for symmetric matrices. Under some constraints 
it is quadratic with logarithmic corrections. We have given an explicit formula 
for the level spacing distribution function in form of a threefold integral.
Then we have considered symmetric matrices with general, other than Gaussian statistics 
for both the diagonal and the non-diagonal elements. We have shown that
the level repulsion again is always linear provided the matrix element
distribution functions are regular at zero value with finite, non-zero weight. We 
have explicitly considered the box-type (uniform), the Cauchy-Lorentz, the 
exponential times singular power law at zero, and the purely exponential matrix element 
distributions. Explicit closed form results for $P(S)$ have been obtained,
except for the singular times exponential tail distributions, although
we could present a very good understanding of the overall behaviour of $P(S)$ also in this case.
 
Our approach can obviously be extended to general 2D complex
random matrices, first to Hermitian complex matrices, where
similar general formulae like equation (\ref{SymmetricP(S)polar})
can be obtained, by using spherical coordinates. Indeed,
it becomes obvious that we shall always have quadratic level repulsion
as long as the matrix element distributions $g(x)$ are regular at zero $x$,
and a formula analogous to (\ref{Symmetriclevelrepulsion}) will apply.
Denoting the diagonal elements by $\pm a$, and the
off-diagonal elements (complex conjugate) by $b\pm i~c$, we obtain

\be \label{Hermitian}
P(S) = \frac{S^2}{8} \int\int \cos \theta ~d\theta~d\varphi ~
g_a(\frac{S}{2} \sin \theta)~ g_b(\frac{S}{2} \cos \theta \sin \varphi)~
g_c(\frac{S}{2} \cos \theta \cos \varphi), 
\ee
as claimed. The details of further analysis will be published in
a separate paper.

We have also discussed the connection between the singular distributions of the matrix elements with
a power law at zero value and the sparsed matrices which describe nearly integrable
systems. In such systems one derives a fractional exponent power law level
repulsion well known phenomenologically but poorly understood theoretically.
Our approach and results promise new advances in this important direction of
research in quantum chaos of mixed type systems initiated in
\cite{Rob1984} and general random matrix theory.

\vspace{0.3in}


\section*{Acknowledgements}

This work was supported by the Cooperation Program between the Universities of
Marburg, Germany, and Maribor, Slovenia, by the Ministry of Higher Education, 
Science 
and Technology of the Republic of Slovenia, by the Nova Kreditna Banka Maribor 
and by TELEKOM Slovenije. We thank Professor Bruno Eckhardt for useful
comments, and Professor Hans-J\"urgen Sommers for communicating to
us his result \cite{Somm}.

\end{document}